# Low-tech solutions for the COVID-19 supply chain crisis

Andrea M. Armani[1,3*], Darrell E. Hurt[2], Darryl Hwang[3,4], Meghan C. McCarthy[2], Alexis Scholtz[3]

[1]Department of Chemical Engineering and Materials Science, University of Southern California, USA [2]Bioinformatics and Computational Biosciences Branch, Office of Cyber Infrastructure and Computational Biology, National Institute of Allergy and Infectious Diseases, National Institutes of Health, Bethesda, MD 20892, USA [3]Department of Biomedical Engineering, University of Southern California, USA, [4]Department of Radiology, University of Southern California, USA

Email: Andrea Armani <armani@usc.edu>

*A global effort is ongoing in the scientific community and in the Maker Movement, which focuses on creating devices and tinkering with them, to reverse-engineer commercial medical equipment and get it to healthcare workers. For these 'low-tech' solutions to have a real impact, it is important for them to coalesce around approved designs*.

Since the first cases were reported in December 2019, the novel coronavirus disease 2019 (COVID-19) has swept across the globe, straining healthcare facilities through sheer case numbers. The World Health Organization declared it a global pandemic on March 11, 2020. Among other symptoms, COVID-19 causes fever, cough, and shortness of breath that can vary from mild to severe, requiring hospitalization and ventilation for the most critical cases.

Although many of the basic symptoms are similar to those of a common cold, COVID-19 is notable for its highly infectious nature and its aggressiveness. For example, on March 17, 2020, the U.S. reached the mark of 100 deaths due to COVID-19. Less than one month later, 26,000 people have died, and over 9,000 healthcare workers have been infected. Similar trends have been observed globally.

As hospitals and healthcare clinics increasingly test and treat patients with COVID-19, healthcare workers expose themselves to the virus at much higher rates than the average person, because they are unable to observe social distancing procedures and other potential methods of mitigating risk while carrying out their jobs. It is therefore of vital importance that healthcare workers have proper personal protection equipment (PPE) not only to prevent further transmission, but also to prevent further strain to the healthcare system that may occur if these clinicians are unable to work due to illness.

However, due to the pandemic's effect on the global supply chain, the stockpiles of PPE are dwindling in many regions, and some hospitals and clinics have gotten so desperate that single-use items are now reused repeatedly. This crisis is not a new problem: supply chain vulnerabilities were exposed during the 2009 H1N1 influenza and 2014 Ebola virus epidemics (1). These vulnerabilities create a critical need for alternative sources of PPE.

This need was quickly recognized by the members of the growing Maker Movement, a global community focusing on 'learning through doing'. As the cost of manufacturing equipment such as 3D printers and electronic components has dropped in recent years, this movement has permeated both formal educational settings and at-home hobbyist circles. Thus, in essence, this movement formed an extremely distributed and agile global network of manufacturers with widely varying capabilities. This network is a naturally occurring component of the maker culture. During the course of this pandemic, the members have focused on tackling three key areas: worker protection, disinfection, and healthcare devices. Their success is due, in large part, to an existing ecosystem that was established prior to COVID-19.

## [H1] The manufacturing ecosystem

Innovative Makers and hobbyists are stepping in to fill the gaps in the PPE supply chain resulting from the COVID-19 public health emergency, and the U.S. Food and Drug Administration (FDA) has issued emergency use authorizations (https://www.fda.gov/medical-devices/emergency-situations-medical-devices/emergency-use-authorizations#covid19ppe) to waive requirements for labeling and good manufacturing practices. Companies and organizations have provided collated collections of designs and challenges to encourage creative solutions and their sharing. Notable examples include Thingiverse (https://www.thingiverse.com/), Matter Hackers (https://www.matterhackers.com/covid-19), and Open Source Medical Supplies (https://opensourcemedicalsupplies.org/), which host a multitude of device designs that have been shared across the web. Organized efforts have also formed through social media and collaboration platforms. The impact overall is positive, but designs are quickly evolving. In many cases, they lack sufficient instructions to inform proper fabrication and use.

Even if the supply chain system is not prepared as a whole, we can capitalize on production methods that allow us to rapidly shift manufacturing to PPE and related supplies. 3D printing technology is well-suited to do this, because it requires little to no modifications to switch from creating one product to the next. However, the agility inherent to 3D printing technologies means that the same input file can be the starting point for almost infinite variations in structural and material composition. This level of variability is detrimental and becomes a risk factor when the product is used as a barrier to an infectious pathogen. Thus, our ability to rapidly respond to the current supply chain crisis — whether through commercial manufacturers or individual Makers — is dependent on determining not just what we should be producing and for whom, but also *how* we should produce it.

The National institutes of health 3D Print Exchange (NIH 3DPX) is a free resource that serves as an open repository of web-based tools for finding, sharing, and creating 3D-printable models related to bioscience and medicine (2). The project was initiated in 2013, during the early days of consumer 3D printing, when existing 3D model repositories lacked biological relevance and accuracy.

The NIH 3DPX COVID-19 Supply Chain Response (https://3dprint.nih.gov/collections/covid-19-response) collection aims to establish standards for both industry and the community. The project is a collaboration among NIH and the National Institute of Allergy and Infectious Diseases, the FDA, the Veterans Health Administration, and America Makes, and emerges from a shared goal of enabling a rapid and safe response to the PPE supply chain through open-source solutions. The objective is to collate and review open-source PPE designs through a systematic and transparent process, resulting in a curated collection of designs that have been vetted and are recommended for community use or in a clinical setting, and designate if a device must have FDA approval or adhere to other standards for manufacturing.

### [H1] Personal protective equipment

The Maker community has focused efforts on making two types of PPE: barrier PPE, such as face shields, and filtering PPE, such as face masks. The success in the fabrication of barrier PPE has been widespread. Face shields are particularly amenable to fabrication using 3D printing due to their simple design, and several variants of face shield designs created in collaboration with medical professionals are available on the NIH 3DPX. Initial designs were launched by Prusa (https://www.prusa3d.com/covid19/), with subsequent versions designed by Budmen in collaboration with Columbia University (https://studio.cul.columbia.edu/face-shield/), among others. In addition, to address the lack of widespread access to 3D printers, single-use face shields, such as the badger shield from the University of Wisconsin (https://making.engr.wisc.edu/shield/), are made from foam and elastic. These types of open-source designs made from accessible materials have unified the global Maker community.

With this immediate success, Makers turned their attention towards respirators. Unfortunately, success in this domain has been limited. Often labeled erroneously as N95 masks, although their quality is not high enough for them to be categorized as such, these home-made masks are 3D-printed models with an integrated filter medium. The quality of the masks can be assessed based on two factors: the fit to the user's face and the type of material used as the filter. In most of these masks, the role of the 3D component is creating an air-tight seal between the airway of the user and the filter material. As such, all of these models are judged based on their ability to conform to the user's face to create a seal, their ability to secure the filter material, and their impedance to air flow.

Makers have created numerous models to meet these challenges, including combining components from different models to address issues such as ease of printing, differing face geometries, and filter availability. The challenge of cushioning the rigid plastic to the relative softness of the human face has been addressed using various materials, such as foam or weather stripping, and even by multi-material printing.

The majority of filters have been commercially available, offering modified versions for use with 3D mask models. Some mask designs include mounting hardware to secure existing filter modules from name brand respirators. Others have adopted the use of furnace or high-efficiency particulate air (HEPA) filters with known ratings; however, there are risks, as many of these filters contain fiberglass. Sourcing N95-grade filtering material is challenging, and it has been a limiting factor to the adoption of 3D-printed filtered masks. With these limitations, clinical use of current 3D-printed mask designs seems unlikely.

# [H1] Disinfection

Disinfection plays a key role in the safety and well-being of healthcare workers and broader society. The conventional disinfection strategy includes a chemical treatment to remove any gross contaminants and then a secondary thermal, chemical vapor, or irradiation treatment to remove any remaining microscopic or nanoscopic materials. However, the conventional auto-clave-style thermal treatment degrades some of the less robust plastic materials, and can degrade fiber-based structures. This limit forces medical facilities to rely on irradiation and chemical vapor methods, which are newer techniques and are not commonly found in smaller clinics. Therefore, to increase the total amount of PPE available to healthcare workers, alternative PPE disinfection systems are needed.

The primary limitation for the Maker community in building a disinfection method is the acquisition of materials. For example, one approach demonstrated by a team from Cleveland Clinic and Case Western Reserve University relied on re-purposing the disinfection capabilities based on ultraviolet-C (UV-C) light in biosafety cabinets scattered in dormant research labs (3). Although successful and quick to implement, this approach relies on existing infrastructure. To overcome this limitation, an alternative strategy focused on distilling the conventional industrial UV-C system to its basic elements.

In such a simplified system, the interior of a plastic bin is spray-painted with a reflective coating, and a conventional UV-C bulb is mounted on the side. Thus, through judicious choice of source intensity and exposure duration, similar performance to that of commercial systems can be achieved (4). The plastic tubs are lightweight and portable, but the throughput is moderate because only a few masks can be disinfected at once. Additionally, although UV-C is ideal for the disinfection of plastic structures such as face shields, there are currently conflicting reports on its suitability for fiber-based materials.

# [H1] Emerging technologies

Ventilators have captured the attention of Makers worldwide since early waves of the pandemic flooded Italy, and it became clear that hospitals lacked the needed quantities of not just PPE, but also equipment. Although regulatory agencies such as the FDA have been reluctant to approve any engineered designs due to the high risks of the life-or-death situations in which ventilators are required, Maker efforts have continued in full force. Given that the COVID-19 outbreak extends across the globe, sharing designs online makes them available in countries not subject to the same restrictions, and where the need for life-saving equipment may be more desperate due to a lack of resources.

Teams of Makers of varying background and experience, from students to veteran engineers, have coalesced at various institutions around the globe to innovate and create emergency ventilators. Robert L. Read, the founder of the nonprofit initiative Public Invention (https://github.com/PubInv/covid19-vent-list), has compiled an extensive repository of resources for open-source ventilators, including analysis and websites of over 80 projects. True to the principles of the Maker community, many of the projects are providing frequent updates and open-source designs with documentation.

The most established design currently is the E-Vent from MIT (https://e-vent.mit.edu/). This E-Vent, first presented in 2010, automates manual resuscitators. Other highly developed and tested designs include that of the AmboVent initiative (https://1nn0v8ter.rocks/AmboVent-1690-108), created by a team in Israel, and of the Open Source Ventilator Project (https://simulation.health.ufl.edu/technology-development/open-source-ventilator-project/) from the University of Florida. Perhaps a sign of the inspiring forces generated by the Maker Movement, in a drastic step away from the traditional trade secrets of industry, Medtronic released full design schematics for its ventilator.

The design of ventilators requires close collaboration with clinicians and may not be an accessible undertaking for many Makers. However, the community has found many other ways to contribute to the COVID-19 pandemic. These contributions are reflected in the increased diversity of designs submitted to NIH 3DPX meant to help relieve stress on the ears of healthcare workers due to wearing a mask all day, and the emergence of designs for hands-free door handles such as those released by Materialise (https://www.materialise.com/en/hands-free-door-opener). Many of these designs embrace the philosophy of the Maker community that anyone can contribute, and has helped foster a sense of the community coming together in a time of public need.

## [H1] Conclusions

Though not the original motivation for the Maker Movement, the benefit of these community-led efforts to the healthcare community and broader society during the COVID-19 pandemic is undeniable. Makers were able to quickly mobilize by leveraging existing tools for source-code dissemination, accelerating innovation and targeted problem-solving. Notably, the COVID-19 emergency has highlighted the power of the Maker community to make a real and immediate impact. Although the emergency use authorizations issued by the FDA for face shields (https://www.fda.gov/media/136842/download) and for systems developed by industry, such as the Battelle Decontamination System (https://www.fda.gov/media/136529/download), which can disinfect thousands of masks at a time using a vapor-phase hydrogen peroxide, are only effective for the duration of the COVID-19 crisis, this does not diminish the important role of the community as a stopgap in this time of need. In future times of crisis, we can learn from the present to harness the energy, creativity, and generosity of Makers.

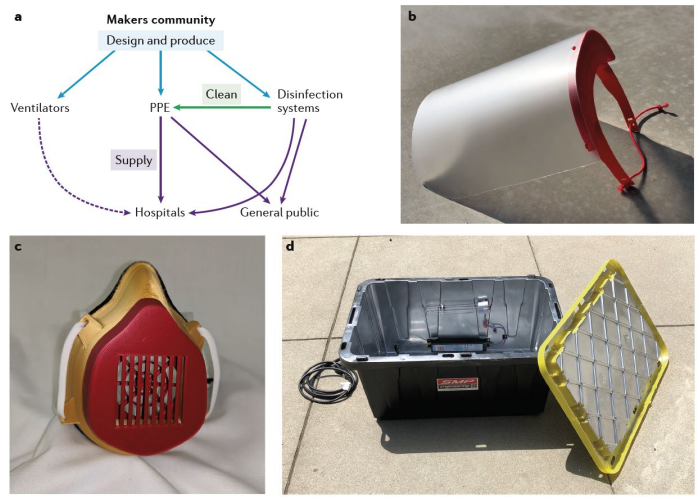

Figure 1| How the Makers are helping. a) Overview of the multi-faceted contributions of the maker community to the COVID-19 pandemic. The dashed arrow indicates a supply line that is not fully established. b) A 3D-printed face shield. c) a 3D-printed face mask. d) A disinfection box using ultraviolet light.

*Acknowledgement*

The authors thank the Army Research Office (W911NF1810033). Current contributions from NIH/NIAID are funded in part through the BCBB Support Services Contract HHSN316201300006W/HHSN27200002.

*Conflict of Interest*

The authors declare no conflict of interest.

Author contributions

All authors contributed equally to the writing of the article.

Webpage links (in order of appearance)

- Emergency Use Authorizations: https://www.fda.gov/medical-devices/emergency-situations-medical-devices/emergency-use-authorizations#covid19ppe

- Thingiverse: https://www.thingiverse.com/

- Matter Hackers: https://www.matterhackers.com/covid-19

- OSMS: https://opensourcemedicalsupplies.org/

- NIH 3DPX: https://3dprint.nih.gov/collections/covid-19-response

- Prusa: https://www.prusa3d.com/covid19/

- Budmen in collaboration with Columbia University: https://studio.cul.columbia.edu/face-shield/

- badger shield from the University of Wisconsin: https://making.engr.wisc.edu/shield/

- CCDS: https://www.battelle.org/inb/battelle-critical-care-decontamination-system-for-covid19

- Public Invention: https://github.com/PubInv/covid19-vent-list

- E-Vent from MIT: https://e-vent.mit.edu/

- AmboVent initiative: https://1nn0v8ter.rocks/AmboVent-1690-108

- Open Source Ventilator Project: https://simulation.health.ufl.edu/technology-development/open-source-ventilator-project/

- Materialise: https://www.materialise.com/en/hands-free-door-opener

- Face shields: https://www.fda.gov/media/136842/download
- Battelle Decontamination System: https://www.fda.gov/media/136529/download